\documentclass[11pt,preprintnumbers,aps,amssymb,nofootinbib,amsmath,superscriptaddress,notitlepage]
{revtex4-1}
\usepackage{epsfig,epsf}
\usepackage{comment}
\usepackage{bm} % puts greek and math symbols in boldface using \bm
\usepackage{color} % {\color{red} ... }
\usepackage{slashed} % Dirac slash
\usepackage{relsize}	%rescalable math 
\usepackage{soul} %spac­ing out (let­terspac­ing), un­der­lin­ing, strik­ing out, etc.
\usepackage{hyperref}
\usepackage{tensor} %\indices{^a_b} produces the properly spaced the tensor indices 
\newcommand{\beq}{\begin{equation}}
\newcommand{\beql}[1]{\begin{equation}\label{#1}}
\newcommand{\eeq}{\end{equation}}
\def\bal#1\gal{\begin{align}#1\end{align}}
\newcommand{\ball}[1]{\bal\label{#1}}
\newcommand{\eq}[1]{(\ref{#1})}
\newcommand{\fig}[1]{Fig.~\ref{#1}}

\DeclareMathOperator{\im}{\mathrm{Im}}

\renewcommand{\b}[1]{{\bm #1}} 
\newcommand{\unit}[1]{\hat {{\bm #1}}} % unit vector

\newcommand{\e}{\varepsilon}

\begin{document}

\title{Spin contribution to the dissociation of bound states in rotating medium in magnetic field}

\author{Kirill Tuchin}

\affiliation{
Department of Physics and Astronomy, Iowa State University, Ames, Iowa, 50011, USA}

\date{\today}

\begin{abstract}

Magneto-rotational dissociation is the decay, by the way of tunneling, of a rotating bound state in the magnetic field. The corresponding probability was recently computed in the quasi-classical approximation using the Imaginary Time Method and was shown to increase with the angular velocity and decrease with the magnetic field strength \cite{Tuchin:2021lxl}. This letter reports the calculation of the quantum correction to the dissociation probability associated with the spin of the tunneling particle. The quasi-classical motion of spin is described by the Bargmann--Michel--Telegdi equation in the rotating frame. It is shown that the spin contribution significantly increases the dissociation probability. Applications to the Quark-Gluon Plasma are touched upon. 

\end{abstract}

\maketitle

%%%%%%%%%%%%%%%%%%%%%%%%%%%%%%%%%%%%%%%%
\section{Introduction}\label{sec:a}

The lifetime of hadrons moving in the magnetic field is reduced due to the Lorentz ionization, which is the dissociation of the bound state due to the electric field in its comoving frame. In the context of the relativistic heavy-ion collisions this effect was analyzed in \cite{Marasinghe:2011bt}. Since then it was realized that the produced nuclear matter possesses large vorticity, which on average points in the same direction as the magnetic field \cite{Csernai:2013bqa,Csernai:2014ywa,Becattini:2015ska,Deng:2016gyh,Jiang:2016woz,Kolomeitsev:2018svb,Deng:2020ygd,Xia:2018tes}. The corresponding magneto-rotational dissociation was recently computed in \cite{Tuchin:2021lxl} at the leading order in $\hbar$ (quasi-classical approximation) and $1/c$ (non-relativistic approximation). The present paper computes the order $\hbar/c$ correction, which describes interaction of the electromagnetic field with the quark spin. The results reported herein apply to bound states in any rotating medium in the magnetic field. However, for the sake of clarity I will refer to hadrons in the Quark-Gluon Plasma. 

Consider a medium rotating with constant angular  velocity $\b \Omega$ in constant magnetic field $\b B$ pointing in the same direction. Let a  hadron traverse this medium with the relative translation velocity $\b V$ as it is dragged along to rotate with the same angular velocity $\b\Omega$. In the hadron comoving frame, the bound state is subject to the mutually orthogonal and constant electric and magnetic fields, the former is given by  $\b E=\b V\times \b B$ while the latter is invariant under the Galilean transformations. Under the action of the electric field, there is a  finite probability for a quark to tunnel through the hadron's potential barrier. It can be calculated in the quasi-classical approximation using the Imaginary Time Method \cite{popov-review}. In this method the dissociation probability $w$ is related to the imaginary part of the restricted action $W$ along the extremal sub-barrier trajectory of a quark escaping the bound state. The extremal trajectory is a solution to the classical equations of motion and as such does not depend on quark's spin. The spin contribution $S'$ emerges as the correction of order $\hbar$ to the classical quark action. Altogether, the dissociation probability reads 
\ball{a2}
w= \exp\{-2\im (W+S')/\hbar\}\,.
\gal
The classical term $W$ was computed in the preceding article  \cite{Tuchin:2021lxl}. This paper computes the spin contribution given by \cite{Marinov:1972nx}
\ball{aa1}
S'= \frac{\mu}{c}\int \tilde F^{\mu\nu}u_\mu a_\nu ds\,,
\gal
where $\mu = e\hbar/2mc$ is quark's magnetic moment, $u^\mu$ its 4-velocity, $\tilde F^{\mu\nu}= \frac{1}{2}\epsilon^{\alpha\beta\mu\nu}F_{\alpha\beta}$ is the dual field tensor, $s$ is the invariant interval $ds= cdt\sqrt{1-v^2/c^2}$ and $a^\mu$ is the Pauli-Lubanski axial 4-vector.  In the instantaneously comoving inertial frame $a^\nu=(0,\b \zeta)$, where  $\b \zeta$  is the double expectation value of the quark spin operator.
In the limit $v\ll c$ and $\Omega r\ll c$, Eq.~\eq{aa1} reduces to the familiar non-relativistic expression 
\ball{aa6}
S'= \mu\int \b B\cdot \b \zeta\, dt+\mathcal{O}(1/c^2)\,.
\gal
The time-dependence of $a$ is governed by the Bargmann--Michel--Telegdi (BMT) equation \eq{f20} \cite{Bargmann:1959gz}. 

The two-body problem in magnetic field in rotating frame can be solved in phenomenologically important and instructive case of a meson made up of a heavy and a light quarks (e.g.\ $D$-meson). This allows the separation of  meson's center-of-mass motion up to the corrections proportional to the ratio of the light to heavy quark masses. Furthermore, the meson's binding potential is assumed to be short-range, i.e.\ the hadron radius is much smaller than the radius of the lowest Landau orbit. In the presence of the long-range interaction one would have to care about the continuity of the quasi-classical wave functions, which contributes only a slowly varying pre-exponential factor in \eq{a2}.   

In the following sections we solve the equation of motion and the BMT equation to obtain $u^\mu(t)$ and $a^\mu(t)$  respectively. These are then used to compute the action \eq{aa1}. The main result is Eq.~\eq{g3} and Figs.~\ref{fig:logW} and \ref{fig:W}.

%%%%%%%%%%%%%%%%%%%%%%%%%%%%%%%%%%%%%%%%
\section{Equations of motion}\label{sec:b}

Equation of motion of a quark  of mass $m$ and electric charge $e$ in the electromagnetic field reads: 
\ball{c2}
\frac{du^\mu}{ds}+\Gamma\indices{^\mu _\nu _\lambda}u^\nu u^\lambda= \frac{e}{mc^2}F^{\mu\nu}u_\nu\,,
\gal
where $u^\mu = dx^\mu/ds$ is 4-velocity. In a frame  rotating with angular velocity $\b \Omega= \Omega \unit z$ with respect to the laboratory frame (which is the center-of-mass frame of the heavy-ion collision)  the metric tensor reads
\ball{b8}
g_{\mu\nu}= \left(\begin{array}{cccc}
1-\Omega^2(x^2+y^2)/c^2& y\Omega /c & -x\Omega /c& 0 \\
y\Omega /c & -1 & 0 & 0 \\ 
-x \Omega/c & 0 & -1 & 0 \\
0 & 0 & 0 & -1
\end{array}\right)\,.
\gal
The corresponding non-vanishing Christoffel symbols are
\ball{c4}
\Gamma\indices{^x _t _t}=-x\Omega^2/c^2\,,\qquad \Gamma\indices{^x _t _y}= \Gamma\indices{^x _y_t}=-\Omega/c\,,\qquad 
\Gamma\indices{^y _t _t}=-y\Omega^2/c^2\,,\qquad \Gamma\indices{^y _t _x}= \Gamma\indices{^y _x_t}=\Omega/c\,.
\gal

Let $\b B= B\unit z$ and $\b E= E\unit y$ be the magnetic and electric fields in the hadron comoving frame. We assume that $\Omega, E, B$ are positive. However, the charge $e$ can be either positive or negative.  The field strength tensor reads
\ball{b10}
F_{\mu\nu}= \left(\begin{array}{cccc}0 & 0& E& 0 \\ 0 & 0 & -B  & 0 \\-E & B & 0 & 0 \\0 & 0 & 0 & 0\end{array}\right)\,.
\gal
Since $\sqrt{-g}=1$ in the rotating frame, the Levi-Civita symbol  $\e^{\alpha\beta\mu\nu}$ is the same as in the Cartesian coordinates.  The dual field strength is given by
 \ball{b11}
 \tilde F^{\mu\nu}=\frac{1}{2}\e^{\alpha\beta\mu\nu}F_{\alpha\beta}=
 \left(\begin{array}{cccc}0 & 0& 0& -B \\ 0 & 0 & 0  & -E \\ 0& 0 & 0 & 0 \\ B & E & 0 & 0\end{array}\right)\,.
\gal
Raising the indices of $F_{\mu\nu}$ by means of the contravariant tensor $g^{\mu\nu}\approx g_{\mu\nu}$ yields 
\ball{b13}
F^{\mu\nu}= \left(\begin{array}{cccc}0 & x\Omega B/c & -E+y\Omega B /c& 0 \\-x\Omega B/c & 0 & -B -y \Omega E/c & 0 \\E-y\Omega B/c & B+y\Omega E/c & 0 & 0 \\0 & 0 & 0 & 0\end{array}\right)+\mathcal{O}(1/c^2)\,.
\gal

The covariant components of the four-velocity  are %Notes p.81
\ball{b15}
%u_t\approx u^t\approx 1\,,\quad  u_x\approx (-v^x+y\Omega)/c\,,\quad  u_y\approx (-v^y-x\Omega)/c\,,\quad u_z= -v^z/c\,,
%\gal
%or, equivalently,
%\ball{b17}
u_\mu = \left(1, -\b v/c- \b \Omega \times \b r/c\right)\,.
\gal
Substituting \eq{b13},\eq{b15} and $s\approx c t$ into \eq{c2} we obtain the equations of motion in the hadron comoving frame at the leading non-relativistic order 
\bal
&\dot v^x-2\Omega v^y-x\Omega^2= \frac{eB}{mc}v^y \,, \label{c7}\\
&\dot v^y+2\Omega v^x-y\Omega^2=\frac{eE}{m}-\frac{eB}{mc}v^x\,,\label{c8}\\
&\dot v^z=0\,.\label{c9}
\gal
where the dotted symbols are time-derivatives. The classical sub-barrier  trajectory that minimizes the restricted action  is a solution to Eqs.~\eq{c7}-\eq{c9}. It can be thought of as the classical motion in the imaginary time \cite{Landau:1991wop}.  It  starts at the imaginary time $t_0$, ends at $t=0$ and satisfies the following boundary conditions \cite{popov-review}:
\bal
\b r(t_0)&= 0\,,\label{c24}\\
\im \b r(0)&= \im \b v(0)=0\,, \label{c25}\\
\frac{1}{2}m v^2(t_0)&=\e_0-mc^2 = -\e_b\,. \label{c26}
\gal
where $\e_b>0$ is the hadron binding energy. Eq.~\eq{c26} determines the initial time $t_0$ of the sub-barrier motion. It is convenient to introduce notations 
\ball{c19}
\omega_\pm = \Omega+\frac{\omega_B}{2}\pm \sqrt{\Omega \omega_B+\frac{\omega_B^2}{4}}\,,
\gal
and 
\ball{c11}
\omega_E= \frac{eE}{mc}\,,\qquad \omega_B= \frac{eB}{mc}\,.
\gal
The quark trajectory that satisfies the  initial conditions \eq{c24},\eq{c25} reads \cite{Tuchin:2021lxl}
\bal
x(\tau)&=\frac{ic\omega_E}{\Omega^2\sinh[(\omega_+-\omega_-)\tau_0]}\left\{ \sinh(\omega_-\tau_0)\sinh(\omega_+\tau)
-\sinh(\omega_+\tau_0)\sinh(\omega_-\tau)\right\}
 \,, \label{c30}\\
y(\tau)&=\frac{c\omega_E}{\Omega^2}\left\{
-\frac{\sinh(\omega_-\tau_0)\cosh(\omega_+\tau)}{\sinh[(\omega_+-\omega_-)\tau_0]}
+\frac{\sinh(\omega_+\tau_0)\cosh(\omega_-\tau)}{\sinh[(\omega_+-\omega_-)\tau_0]}-1
 \right\}\,,\label{c31}
\gal
and $z(\tau)=0$, where $\tau=it$ is the real Euclidean time such that  $\tau_0\le \tau \le 0$ and $\tau_0<0$.

%%%%%%%%%%%%%%%%%%%%%%%%%%%%%%%%%%%%%%%%
\section{Spin precession}\label{sec:f}

Spin precession is described by the Pauli-Lubanski 4-vector $a^\mu$ which is orthogonal to the quark 4-velocity \cite{Berestetsky:1982aq}
\ball{f2}
a^\mu u_\mu=0\,,
\gal
and is normalized according to 
\ball{f3}
a^\mu a_\mu= -\b \zeta^2\le 1\,.
\gal 
In the pure state $\b \zeta^2 =1$.  The covariant components are obtained using the metric tensor \eq{b8}:
\ball{f5} 
a_\mu=\left(a^0+\b\Omega\cdot (\b a\times \b r)/c, -\b a\right)+\mathcal{O}(1/c^2)\,.
\gal
Employing \eq{f2},\eq{f5} and \eq{b15} we find 
\ball{f7}
a_0= \frac{\b a\cdot \b v}{c}\,,\quad a^0= \frac{\b a}{c}\cdot \left(\b v+\b \Omega\times \b r\right)\,.
\gal
Since $|a_0|\ll |\b a|$ it follows that  $a^\mu a_\mu\approx -\b a^2$ which in turn implies $|\b a|= |\b \zeta|$ in any frame  within the Galilean group. 
%Since in the instantaneously comoving inertial frame $a^\mu=(0,\b \zeta)$, in a frame rotating with respect to it 
%\ball{f9}
%a^\mu= (0,\zeta_x \cos\Omega t +\zeta_y \sin\Omega t, -\zeta_x \sin\Omega t+\zeta_y \cos \Omega t, \zeta_z)\equiv (0,\b \zeta_R) \,.
%\gal
%Vector $\b \zeta_R$ precesses around the $z$-axis with frequency $\Omega$, but its $z$-component remains constant. 

The time-dependence of $a$ in a homogenous external field is governed by the BMT equation \cite{Bargmann:1959gz}:
\ball{f20}
\frac{da^\mu}{ds}+\Gamma\indices{^\mu _\nu _\lambda}a^\nu u^\lambda= \frac{e}{mc^2} F^{\mu\nu}a_\nu\,.
\gal
Substituting \eq{b13} and \eq{f5} and keeping only the leading non-relativistic terms yields
\ball{f22}
\dot a^x&= \left( \omega_B +\Omega\right)a^y\,,\qquad 
\dot a^y= -\left( \omega_B+\Omega\right)a^x\,,\qquad
\dot a^z=0\,.
\gal
The last of Eqs.~\eq{f22} implies that the projection of spin onto the magnetic field/angular velocity direction is a constant of motion. The two other equations can together be written down as an equation for the complex quantity $ a_\bot = a^x+i a^y$:
\ball{f26} 
\dot a_\bot = -i\left( \omega_B+\Omega\right)a_\bot
\gal
whose solution is 
\ball{f28}
a_\bot(t) =  a_\bot(t_0) \exp\left\{
-i(\omega_B+\Omega) (t-t_0)
\right\}\,,
\gal
where $a_\bot(t_0)= a^x_0+ia^y_0$ is a constant. 
In the Euclidean time $\tau=it$ Eq.~\eq{f28} reads
\ball{f33}
a^x(\tau)&= a_0^x\cosh[(\omega_B+\Omega) (\tau-\tau_0)]\,,\qquad 
a^y(\tau)= ia_0^y\sinh[(\omega_B+\Omega) (\tau-\tau_0)]\,.
\gal

%%%%%%%%%%%%%%%%%%%%%%%%%%%%%%%%%%%%%%%%
\section{Dissociation probability}\label{sec:d}

The leading term in the restricted action $W$ appearing in \eq{a2} can be computed using the trajectory \eq{c30},\eq{c31}. The result is  \cite{Tuchin:2021lxl}:
\bal
W=& \int_{t_0}^0(L+\e_0)dt- \b p\cdot \b r|_{t=0} \label{d2}\\
 =&i\frac{m\omega_E^2}{8\Omega^4}\frac{\omega_+-\omega_-}{\sinh^2[(\omega_+-\omega_-)\tau_0]}
\big\{-2\tau_0(\omega_+-\omega_-)+2\tau_0\omega_+\cosh(2\omega_-\tau_0)-2\omega_-\tau_0\cosh(2\omega_+\tau_0)\nonumber
 \\
&-\sinh[2(\omega_+-\omega_-)\tau_0]-\sinh(2\omega_-\tau_0)+\sinh(2\omega_+\tau_0)\big\}\,.\label{d5}
\gal
Eq.~\eq{c26} determines implicitly the initial instant of the sub-barrier motion  $\tau_0$: %Notes p.85
\ball{c34}
\gamma^2= \frac{\omega_B^2}{\Omega^4\sinh^2[(\omega_+-\omega_-)\tau_0]}
\big\{
2\omega_+\omega_-\sinh(\omega_-\tau_0)\sinh(\omega_+\tau_0)\cosh[(\omega_+-\omega_-)\tau_0]
&\nonumber\\
-\omega_+^2\sinh^2(\omega_-\tau_0)-\omega_-^2\sinh^2(\omega_+\tau_0)\big\}
\,,&
\gal
where we introduced a positive dimensionless \emph{adiabaticity} parameter \cite{Popov:1997-A,Popov:1998-A}
\ball{c35}
 \gamma= \sqrt{\frac{2\e_b}{m}}\frac{B}{E}\,.
\gal 
Eliminating $\tau_0$ from \eq{d5} and \eq{c34} and substituting into in \eq{a2} yields the desired dissociation probability. 

The spin contribution  to the action is given by \eq{aa1}. It can be expanded in powers of $1/c$ using  \eq{b11} and \eq{b15} yielding
\bal
S'&= \mu\int \left\{ \b B\cdot \b a+\b E\cdot \left(\frac{\b v}{c}\times \b a\right)+ \frac{\b \Omega}{c}\cdot \left[(\b E\times \b a)\times \b r\right]\right\}dt \label{g1}\\
&=\frac{\hbar a^z}{2}\int \left\{ \omega_B + \frac{\omega_E}{c}(\Omega y -v^x)\right\} dt\,,\label{g2}
\gal
where it was used that $v^z=0$ and $a^z$ is a constant. Substituting \eq{c31} and integrating furnishes the main result
\ball{g3}
S'&= \frac{\hbar a^zi}{2} \left\{\left(\omega_B-\frac{\omega_E^2}{\Omega}\right)\tau_0+ 
\frac{\omega_E^2(\omega_+-\omega_-)}{\Omega^3}\frac{\sinh(\omega_+\tau_0)\sinh(\omega_-\tau_0)}{\sinh[(\omega_+-\omega_-)\tau_0]}\right\}\,.
\gal

It is useful to define the dimensionless parameters introduced in \cite{Tuchin:2021lxl}
\ball{c50}
\eta= -\omega_B\tau_0\,,\quad \kappa=\Omega/\omega_B\,.
\gal 
Note that since $\tau_0<0$, $\eta>0$ for positive charges, and $\eta<0$ negative ones.  In terms of $\eta$ and $\kappa$ 
Eqs.~\eq{d2} and  \eq{d5} read
%% p.92
\bal
W=&\frac{im\omega_E^2\sqrt{4\kappa+1}}{8\kappa^4\omega_B^3\sinh^2[\eta\sqrt{4\kappa+1}]}
\bigg\{ 2\eta \sqrt{4\kappa+1}-\eta\left( 2\kappa+1+\sqrt{4\kappa+1}\right)\cosh\left[\eta\left(2\kappa+1-\sqrt{4\kappa+1}\right)\right]\nonumber\\
&+\eta\left(2\kappa+1-\sqrt{4\kappa+1}\right)\cosh\left[\eta\left(2\kappa+1+\sqrt{4\kappa+1}\right)\right]+\sinh\left[2\eta\sqrt{4\kappa+1}\right]\nonumber\\
&+\sinh\left[\eta\left(2\kappa+1-\sqrt{4\kappa+1}\right)\right]-\sinh\left[\eta\left(2\kappa+1+\sqrt{4\kappa+1}\right)\right] 
\bigg\}\,, \label{c60}
\gal
\bal
\gamma^2=&\frac{1}{4\kappa^4\sinh^2\left[ \eta\sqrt{4\kappa+1}\right]}\bigg\{
8\kappa^2\sinh\left[\eta\left(\kappa+1/2-\sqrt{\kappa+1/4}\right)\right]\nonumber\\
&\times\sinh\left[\eta\left(\kappa+1/2+\sqrt{\kappa+1/4}\right)\right]
\cosh\left[\eta\sqrt{4\kappa+1}\right]\nonumber\\
&-\left(2\kappa+1+\sqrt{4\kappa+1}\right)^2\sinh^2\left[\eta\left(\kappa+1/2-\sqrt{\kappa+1/4}\right)\right]
\nonumber\\
&-\left(2\kappa+1-\sqrt{4\kappa+1}\right)^2\sinh^2\left[\eta\left(\kappa+1/2+\sqrt{\kappa+1/4}\right)\right]
\bigg\}\,. \label{c65}
\gal
Eq.~\eq{g3} can be cast in the form
\ball{g7}
S'= \frac{\hbar a^zi}{2}\left\{ -\eta + \frac{E^2}{B^2}\frac{\eta}{\kappa}-\frac{E^2\sqrt{1+4\kappa}}{2\kappa^3 B^2} 
\frac{\cosh\left[\eta(2\kappa+1)\right]-\cosh\left[\eta\sqrt{4\kappa+1}\right]}{\sinh[\eta\sqrt{4\kappa+1}]}
\right\}\,.
\gal

As explained in Introduction,  the ratio $E/B=V$ is the hadron translation velocity. For a given $V$ increase of the magnetic field $B$ results in decrease of $\im W$, while $\im S'$ does not change. Since the applicability of the quasi-classical approximation hinges on the assumption that the terms proportional to $\hbar$ are small, 
the current method breaks down for very strong magnetic fields. For illustration consider a non-rotating system in strong electric field so that $|\eta|\approx \gamma\ll 1$. In this case
\ball{g10} 
\im W&\approx \frac{m^2}{3|e|E}\left( \frac{2\e_b}{m}\right)^{3/2}\,,\qquad
\im S'\approx -\frac{\hbar}{2c} a^z\frac{B}{E}\left( \frac{2\e_b}{m}\right)^{1/2}\,.
\gal
It follows that the magnetic field must satisfy the condition $B\ll \e_b/\mu$. In relativistic heavy-ion collisions this condition is satisfied. In fact, the maximal field $\e_b/\mu$  is smaller but comparable to the critical Schwinger's value $B_c= m^2c^3/e\hbar$. 

\begin{figure}[ht]
\begin{tabular}{cc}
 \includegraphics[height=5cm]{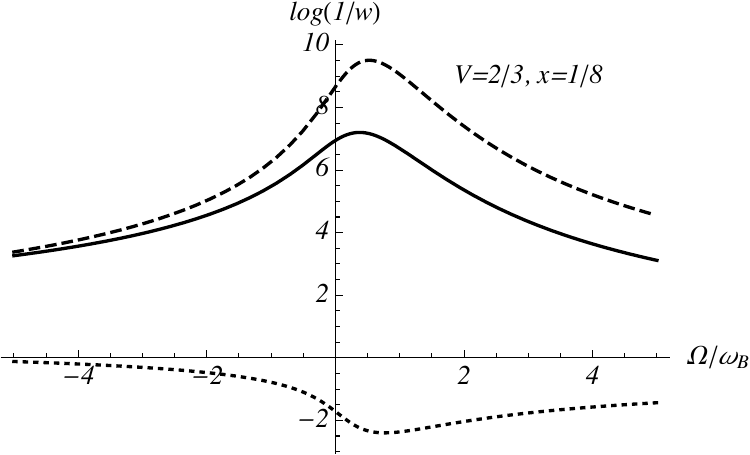} &
          \includegraphics[height=5cm]{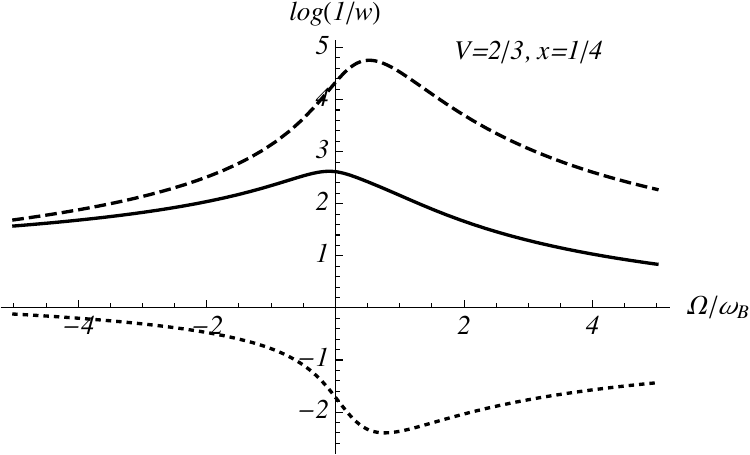}\\
     \includegraphics[height=5cm]{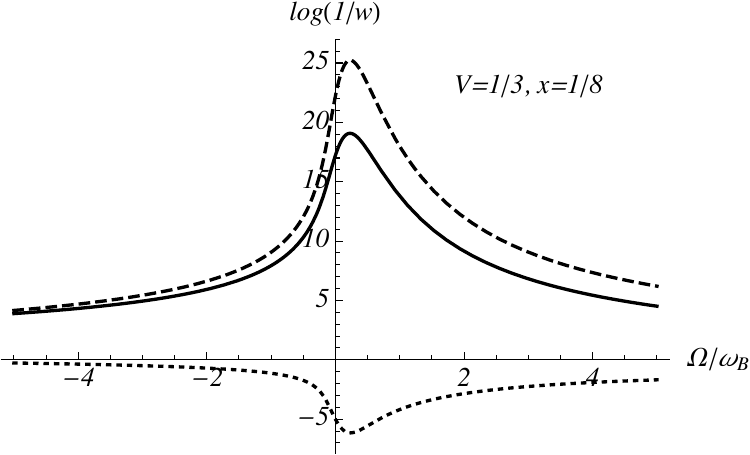} &
          \includegraphics[height=5cm]{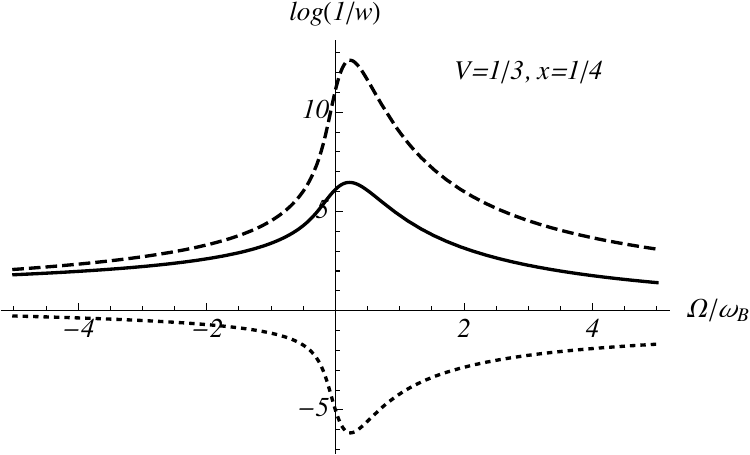} \\
           \includegraphics[height=5cm]{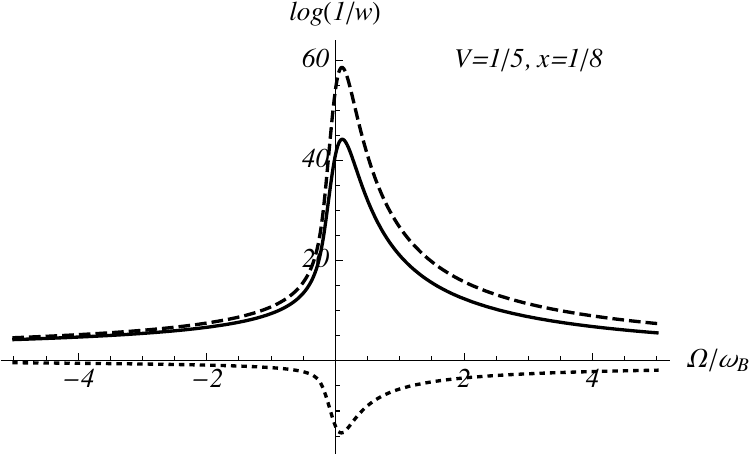} &
          \includegraphics[height=5cm]{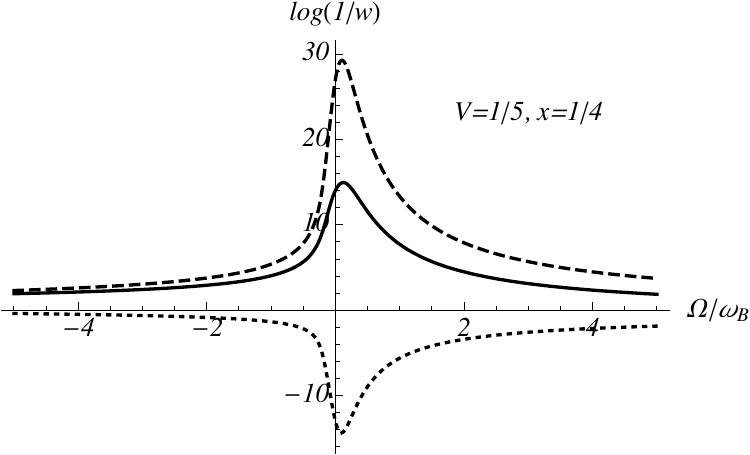} 
                       \end{tabular}
  \caption{$2\im W$ (dashed line), $2\im S'$ (dotted line), and their sum (solid line) as functions of $\Omega/\omega_B$ at different values of  $V=E/B$ and $x=|e|B/m^2$. The binding energy is set at  $\e_b=mc^2/2$, hence $\gamma=1/V$, the quark is assumed to be initially polarized along the $z$-direction: $a^z=1$. }
\label{fig:logW}
\end{figure}

\begin{figure}[ht]
\begin{tabular}{cc}
\includegraphics[height=5cm]{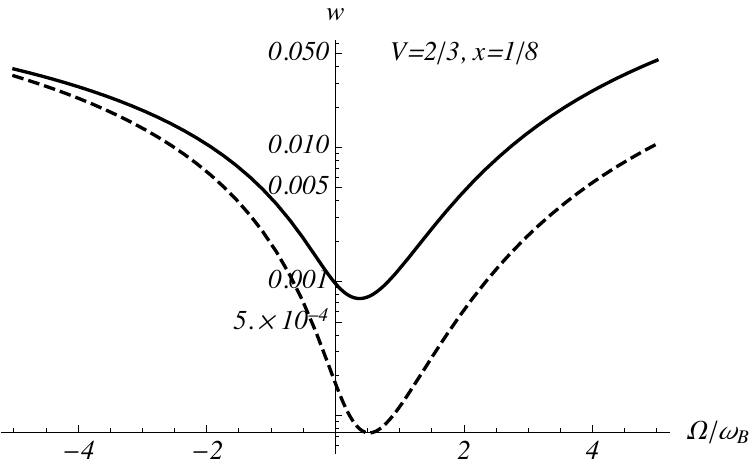} &
          \includegraphics[height=5cm]{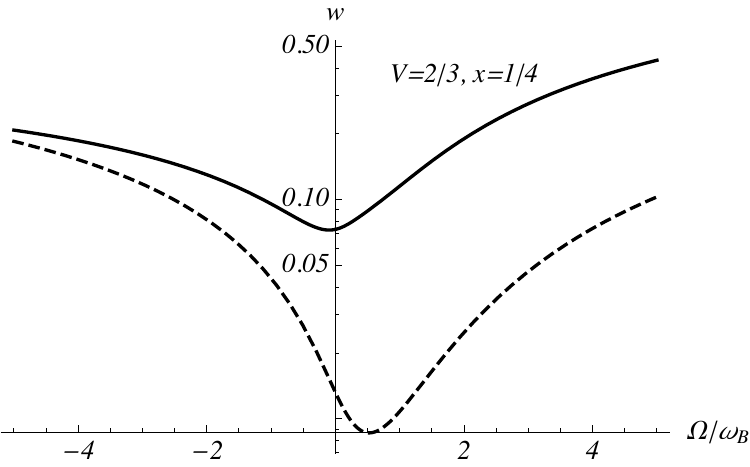}\\
     \includegraphics[height=5cm]{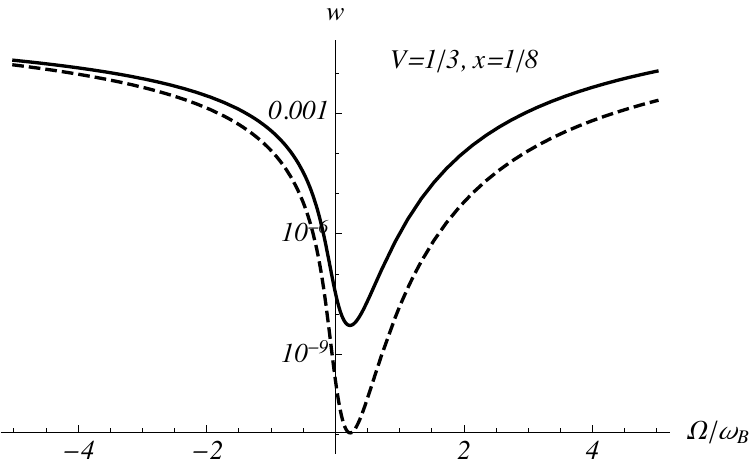} &
          \includegraphics[height=5cm]{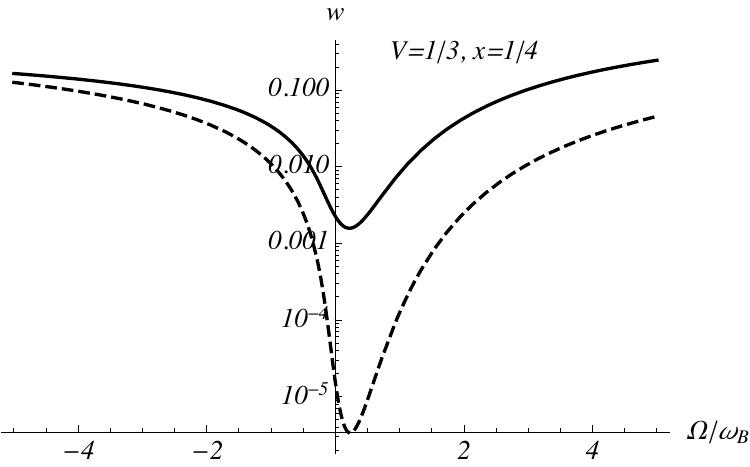} \\
           \includegraphics[height=5cm]{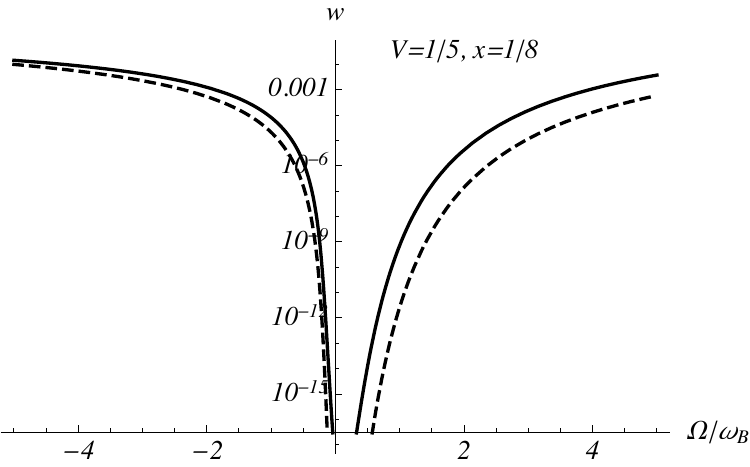} &
          \includegraphics[height=5cm]{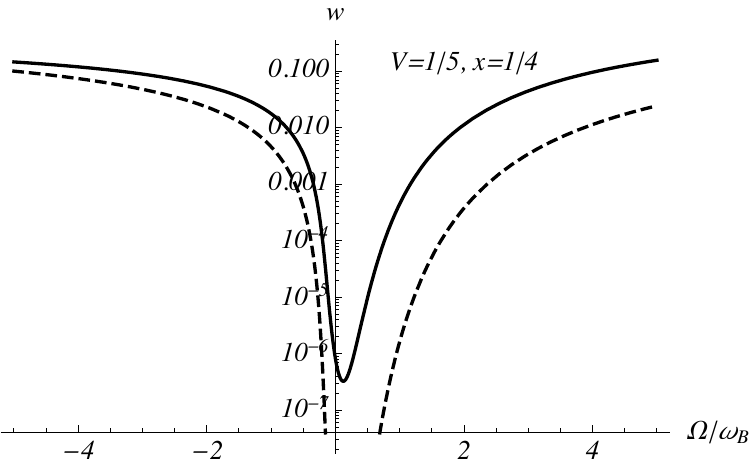} 
             \end{tabular}
  \caption{Dissociation probability $w$ with (solid line) and without (dashed line) the spin term as functions of $\Omega/\omega_B$ at different values of  $V=E/B$ and $x=|e|B/m^2$. The binding energy is set at  $\e_b=mc^2/2$, hence $\gamma=1/V$, the quark is assumed to be initially polarized along the $z$-direction: $a^z=1$. }
\label{fig:W}
\end{figure}

Functions $2\im W$, $2\im S'$ and their sum are plotted in \fig{fig:logW}. It is seen that the spin contribution is negative implying that it increases the dissociation probability. Moreover, since it appears in the exponent, even a small correction produces large change in the total probability as one can see on \fig{fig:W}. \fig{fig:logW} clearly shows that the non-relativistic  quasi-classical approximation improves as $V$ and $B$ decrease.

%%%%
%%%%%

%%%%
%%%%%

%%%%%%%%%%%%%%%%%%%%%%%%%%%%%%%%%%%%%%%%
\section{Discussion}\label{sec:s}

The spin contribution to the magneto-rotational dissociation is the leading quantum relativistic correction to the quasi-classical dissociation probability.  Fig.~\eq{fig:W} indicates that it significantly enhances the magnitude of the effect. However, the qualitative features of the magneto-rotational dissociation, discussed in detail in the preceding paper on this subject \cite{Tuchin:2021lxl}, remain essentially the same. Namely, the dissociation probability increases with the angular velocity due to the centrifugal force and decreases with the magnetic field due to its confining action. One can also observe that the positive charges, corresponding to $\omega_B>0$ have smaller probability to be torn off the hadron than the negative charges with $\omega_B<0$.  

The results obtained in this paper apply to any rotating medium in the magnetic field. Such systems are abound in astro, condensed matter and atomic physics. However, the main motivation of this work has been possible applications to the relativistic heavy-ion phenomenology. Quark-Gluon Plasma (QGP), the matter produced in the relativistic heavy-ion collisions, is known to be rotating and a subject to the magnetic field. Even though $\Omega$ and $B$ vary in space and time it was argued in  \cite{Marasinghe:2011bt} that these variations  can be ignored in the first approximation. This is certainly true for the time-dependence of the magnetic field in light of the recent observation in \cite{Stewart:2021mjz} that $B$ is nearly constant for the most of the QGP history. One can thus use the results of this paper to estimate, at least qualitatively, the probability of  the magneto-rotational dissociation for a given magnetic field strength and vorticity. It is estimated that $|\omega_B|$ is of the same order of magnitude as $\Omega$ at the collision energy of  $\sqrt{s}\sim 50-130$~GeV \cite{Deng:2012pc,Deng:2016gyh,Deng:2020ygd}. Fig.~\eq{fig:W} indicates that the dissociation is strongest at $\Omega\gg |\omega_B|$ which occurs at lower collision energies. The results derived in this work can be used for a qualitative discussion of the general features of the vorticity dependence of the dissociation probability of heavy hadrons. This provides a benchmark for the future phenomenology of the magneto-rotational effect.

%%%%%%%%%%%%%%%%%%%%%%%%%%%%%%%%
\acknowledgments

This work  was supported in part by the U.S.\ Department of Energy under Grant No.\ DE-FG02-87ER40371.

%% appendix 
%\appendix
%\section{}\label{appA}

%%%%%%%%%%%%%%%%%%%%%%%%%%%%%%%%%%%%%

\end{document}